\documentclass{vldb}
\usepackage{graphicx}
\usepackage{balance}  

\usepackage[utf8]{inputenc} 
\usepackage[T1]{fontenc}    
\usepackage{lmodern}
\usepackage{hyperref}       
\usepackage{url}            
\usepackage{booktabs}       
\usepackage{amsfonts}       
\usepackage{nicefrac}       
\usepackage{microtype}      

\usepackage{epstopdf}
\usepackage{subcaption}
\usepackage{algorithm}
\usepackage{algorithmic}
\usepackage{amsmath}

\usepackage{color}
\newtheorem{theorem}{Theorem}
\newtheorem{definition}{Definition}

\begin{document}


\title{Arrays of (locality-sensitive) Count Estimators (ACE): High-Speed Anomaly Detection via Cache Lookups}



%
%
%
%

\numberofauthors{2} 

\author{
%
%
\alignauthor
Chen Luo\\
       \affaddr{Department of Computer Science}\\
       \affaddr{Rice University}\\
       \affaddr{Houston, TX, USA}\\
       \email{cl67@rice.edu}
\alignauthor
Anshumali Shrivastava\\
       \affaddr{Department of Computer Science}\\
       \affaddr{Rice University}\\
       \affaddr{Houston, TX, USA}\\
       \email{anshumali@rice.edu}
}


\maketitle

\begin{abstract}
Anomaly detection is one of the frequent and important subroutines deployed in large-scale data processing systems. Even being a well-studied topic, existing techniques for unsupervised anomaly detection require storing significant amounts of data, which is prohibitive from memory and latency perspective. In the big-data world existing methods fail to address the new set of memory and latency constraints. In this paper, we propose ACE (Arrays of (locality-sensitive) Count Estimators) algorithm that can be 60x faster than the ELKI package~\cite{DBLP:conf/ssd/AchtertBKSZ09}, which has the fastest implementation of the unsupervised anomaly detection algorithms. ACE algorithm requires less than $4MB$ memory, to dynamically compress the full data information into a set of count arrays. These tiny $4MB$ arrays of counts are sufficient for unsupervised anomaly detection. At the core of the ACE algorithm, there is a novel statistical estimator which is derived from the sampling view of Locality Sensitive Hashing(LSH). This view is significantly different and efficient than the widely popular view of LSH for near-neighbor search. We show the superiority of ACE algorithm over 11 popular baselines on 3 benchmark datasets, including the KDD-Cup99 data which is the largest available benchmark comprising of more than half a million entries with ground truth anomaly labels.
\end{abstract}

\section{Introduction}
The problem of Anomaly (or outlier) detections is the task of identifying instances (or patterns) in data that do not conform to the expected behavior~\cite{chandola2009anomaly}. These non-conforming examples are popularly referred to as anomalies, or outliers, sometimes interchangeably. Anomaly (or Outlier) detection algorithms are frequently required in large data processing applications. Anomaly detection is one of the very well studied topics due to a variety of domain applications which include fraud detection, unusual activity detection in video surveillance, intrusion detection in network traffic, etc.

Anomaly detection can be either supervised~\cite{leung2005unsupervised} or unsupervised~\cite{gornitz2013toward}. Supervised anomaly detection leverages machine learning algorithms, such as classification, over datasets labeled as anomalous/non-anomalous. However, there are three major issues with supervised anomaly detection algorithms: 1) In most applications, label information about anomalies is not available, 2) Anomalies are rare, and hence there is a huge class imbalance, and 3) Supervised algorithms needs to be re-trained for drifting data distributions with new label information. Drifting data distribution is quite common in big-data systems, where supervised re-training is prohibitive. Therefore, we are interested in unsupervised anomaly detection which does not require any label information, and which can automatically deal with changes in data distributions over time. We briefly describe some of the modern challenges in unsupervised anomaly detection that we will address in this work.\\

{\bf Challenge 1: High-Speed Drifting Data} Many streaming applications demand fast-response and real-time inference from dynamic and drifting high volumes of sensor data over time. Most anomaly detection applications, for example over the web-network servers, requires dealing with unprecedented amounts of data in a fraction of seconds. The data distribution is constantly changing, and it is often bursty~\cite{kleinberg2002bursty}.
Detecting anomaly events in real-time, such as DDoS (Distributed Denial of Service) attacks, network failures, etc., is highly beneficial in monitoring network performance degradation and service disruptions.  Network attacks can be very costly to recover and may require a complete restart of the service.  Detecting anomalies over video streams from surveillance cameras is critical for identifying suspicious and terrorist activities before they happen. With the increase in sensor resolutions, the data generated is ultra-high dimensional adding to the difficulty.\\

{\bf Challenge 2: Ultra-Low Memory Budget for Cache Utilization} In many high-speed streaming applications, such as High Energy Physics (HEP) and network servers, algorithms communicating with DRAM (main memory) is just too slow for the rate at which the data is generated. Ideally, we need ultra-low memory anomaly detection algorithm that can entirely sit in the cache which can be around 2-10 times faster than accessing main memory~\cite{conway2010cache}. However, the cache memory is limited to few Megabytes. Another critical pushing need for ultra-low memory algorithm is anomaly detection on mobile phones or smart sensors. Algorithms which requires significant computation and memory is prohibitive for these low-memory low-power platforms.

Owing to the significance of the above two challenges and the impact of solving them, mining high-speed streaming data is among the top-10 big-data challenges.\\

{\bf Focus:} The focus of this paper is on designing a novel anomaly detection algorithm that can deal with the above two challenges.\\

{\bf Popular approaches for Unsupervised Anomaly Detection:} Existing methods for unsupervised anomaly detection usually calculate some statistics (or score) between the data $\mathcal{D}$ and the given point of interest $x$, given by $S(x,\mathcal{D})$. The expected value of this statistics for non-outlier points is different from that of the outlier point. The deviation of this statistics, for a given query point of interest $q$, is a measure of the anomalous behavior of $q$. If the deviation is found to be significant, the point $q$ is reported as an anomaly or outlier.

There are numerous approaches for unsupervised anomaly detection in literature. We review and compare with 11 of these popular methods in our experiments. See section~\ref{sec:baseline} for brief descriptions of several popular algorithms.   Unsupervised anomaly detection can be broadly categorized into two categories: 1) Near-Neighbor (NN) Based and 2) Aggregate Statistics Based. NN based approaches typically define the outlier score of a point $q$ based on the difference between $q$'s own behavior compared to the behavior of $q$'s near-neighbors. The first category is the most common category. There are several implementations available. A notable among them is the ELKI package~\cite{DBLP:conf/ssd/AchtertBKSZ09} which is currently one of the most popular packages for outlier detection because of highly optimized implementations and usage of smart near-neighbor search algorithms.

Aggregate statistics based methods, on the other hand, define the outlier score of a point $q$ based on the expected behavior of a global function $S(q,\mathcal{D})$ of the data $\mathcal{D}$, relative to $q$.  A notable method among them is ABOD(Angle-based outlier detection)~\cite{pham2012near}. ABOD computes the variance of the angle formed by different pairs of points, in the dataset, incident on the point of interest $q$.   It is expected that the outlier will have a negligible variance.  See~\cite{pham2012near} for details.

Both the categories of anomaly detection algorithms require storing the complete dataset to either compute near-neighbor or the desired statistics from the dataset. The bottleneck computational cost is at least one pass over the data to either calculate the near-neighbor or the statistics. Thus, these methods have poor computational and memory requirements. Furthermore, change in the distribution of data requires storing and processing larger set of observations. Therefore, the performance of standard algorithms gets worse with increasing data. \\

{\bf Sampling and Fast Near-Neighbors:} To work around the computational requirements it is natural to resort to fast alternatives~\cite{chakrabarti2015bayesian}. There are plenty of techniques which exploits efficient near-neighbor capabilities to speed up NN. However, they still require storing the data in the memory. Even with the computational speedups, the methodologies are still slow for ultra-high speed data mining, as an accurate near-neighbor search over large data is still a costly operation.

Relying on random sampling and projections of the data to estimate the aggregate statistics efficiently is not new~\cite{chandola2009anomaly}.  For example, recently, ~\cite{pham2012near} showed that using smart random sampling and hashing algorithms, we can speed up the anomaly detection and also reduce the memory requirement. Instead of storing all the data points, we only need few random samples and their quantized projections. They proposed FastVOA which uses a modified ABOD statistics that can be estimated in near-constat time and is as good as ABOD for anomaly detection.

However, these approximate methods still require storing a significant number of data samples, which makes the algorithm slow. FastVOA involves various computation of medians and other costly statistics.  Our experiments show that the sampling based FastVOA approach is significantly slower than fast NN based alternatives.\\

{\bf Our Contributions:} We propose a family of statistics which provides a ``sweet" spot between the discriminative power of the statistics and the resource efficiency for anomaly detection. These special statistics, due to their form, can be efficiently computed in ultra-low memory and does not require storing even a single data sample. Furthermore, any updates to the data can be incorporated on the fly making our proposal ideal for high-speed data applications.

Our proposed family of statistics are derived from the collision probability of locality sensitive hashing (LSH) functions. We show that these classes of statistics have strong discriminative property for identifying outliers and most importantly,  it can be accuracy estimated using Arrays of Count Estimators (ACE), a novel and tiny LSH based data strucutre. Designing these estimators requires taking the sampling view of LSH rather than the widely popular near-neighbor search view. To the best of our knowledge, this is the first work which uses LSH counts as unbiased estimators of statistics.

We demonstrate, empirically and theoretically, that the proposed LSH based count estimators are significantly more accurate than random sampling approaches.  Our ACE algorithm only requires computing few Locality Sensitive Hashes of the data and a small set of count array lookups to estimate the proposed statistics sharply.  Our approach does not require even a single distance computation.  The theory and the class of estimators presented in the paper,  could of independent interest in itself.

We demonstrate rigorous experimental evidence on three public outlier detection benchmarks including the largest publicly available benchmark dataset KDD-cup99 HTTP dataset having more than half a million labeled instances.  Empirically, our algorithm only requires around $4MB$ of memory and near-constant amount of computations,  irrespective of the size of the datasets. Thus, our algorithm can exploit fast L3 caches (Level 3 caches), which can be significantly faster than dealing with main memory.

We provide a comparison of our algorithm with 11 different methodologies, which include some of the fastest and most popular anomaly detection algorithms. Our experiment shows that we are around at least 60x faster than of the best performing competitor on the largest benchmark KDD-cup99 HTTP dataset. This disruptive speedup is not surprising given the computational simplicity of our algorithm and ultra-low memory print than can leverage Level 3 cache.\\

{\bf Organization:} We cover some basics of locality sensitive hashing (LSH) and signed random projections (SRP) in Section~\ref{sec:background}. We introduce our ``sweet" measure for anomaly detection in Section~\ref{sec:proposal}. We then show why it is a good notion of outlierness in Section~\ref{sec:whydiscrim}. We then propose our ACE algorithm in Section~\ref{sec:ACEAlgo}. We later argue how the simple ACE algorithm and the proposed measure are connected. We show analysis of the ACE estimators and its properties, along with its superiority over random sampling, in Section~\ref{sec:theory}. Experimental comparisons and evaluations are provided in Section~\ref{sec:exp}. We have deferred the proofs of theorems to the appendix, in the end, for better readability.

\section{Background: Locality Sensitive Hashing}
\label{sec:background}
Locality-Sensitive Hashing (LSH) \cite{indyk1998approximate} is a popular technique for efficient approximate nearest-neighbor search. LSH is a family of functions, such that a function uniformly sampled from this hash family has the property that, under the hash mapping, similar points have a high probability of having the same hash value. More precisely,
Consider $\mathcal{H}$ a family of hash functions mapping $\mathbb{R}^D$ to a discrete set $[0,R-1]$.
\begin{definition} \label{def:lsh}{\bf Locality Sensitive Hashing (LSH) Family}\ A family $\mathcal{H}$ is called $(S_0,cS_0,u_1,u_2)$-sensitive if for any two point $x,y \in \mathbb{R}^d$  and $h$ chosen uniformly from $\mathcal{H}$ satisfies the following:
\begin{itemize}
\item if $Sim(x,y)\ge S_0$ then ${Pr}_\mathcal{H}(h(x) = h(y)) \ge u_1$
\item if $ Sim(x,y)\le cS_0$ then ${Pr}_\mathcal{H}(h(x) = h(y)) \le u_2$
\end{itemize}
 \end{definition}
A collision occurs when the hash values for two data vectors are equal, meaning that $h(x) = h(y)$. The probability of a collision for a LSH hash function is generally proportional to some monotonic function of similarity between the two data vectors, i.e., $Pr[h(x) = h(y)] \propto f(\text{sim}(x,y))$, where $\text{sim}(x,y)$ is the similarity under consideration and $f$ is some monotonically increasing function. Essentially, similar items are more likely to collide with each other under LSH mapping.

LSH is a very well studied topic in computer science theory and database literature. There are a number of well know LSH families in the literature. Please refer~\cite{gionis1999similarity} for details. The most popular one is Signed Random Projections~\cite{charikar2002similarity}.

\subsection{Signed Random Projections(SRP)}
\label{sec:srp}
Signed Random Projections(SRP) is an LSH for the cosine similarity measure, which originates from the concept of \emph{\bf randomized rounding(SRP)}~\cite{goemans1994879,charikar2002similarity}. Given a vector $x$, SRP utilizes a random $w$ vector with each component generated from i.i.d. normal, i.e., $w_i \sim N(0,1)$, and only stores the sign of the projection. Formally SRP family is given by
$$h_w(x) = sign(w^Tx).$$
It was shown in the seminal work~\cite{goemans1994879} that collision under SRP satisfies the following equation:
\begin{equation}
    \label{eq:srp}
    Pr_w(h_w(x) = h_w(y)) = 1 - \frac{\theta}{\pi},
\end{equation}
where  $\theta = cos^{-1}\left( \frac{x^Ty}{||x||_2 ||y||_2}\right)$. The term $\frac{x^Ty}{||x||_2 ||y||_2}$, is the cosine similarity.

If we generate $K$ independent SRP bits, by sampling $w$ independently $k$ times, and use the generated $K$-bit number as the hash value, then the collision probability with the new hash function $H$ becomes
\begin{align} \label{eq:metasrp}
Pr(H(x) = H(y))= (1 - \frac{\theta}{\pi})^K
\end{align}
 by the simple multiplicative law of probability. We will be using this observation heavily in our work.

\subsection{Progress in making Locality Sensitive Computations Faster}
\label{sec:FastJL}
Over the last decade, there has been a significant advancement in reducing the amortized computational and memory requirements for computing several LSH signatures of the data vector.  For random projections based LSH, of which signed random projection is a special case, we can calculate $m$ LSH hashes of the data vector, with dimensions $d$, in time $O(d\log{d} + m)$, a significant improvement over $dm$. This speedup is possible due to the theory of Fast-Johnson–Lindenstrauss transformation~\cite{ailon2006approximate,dasgupta2011fast} . On the orthogonal side, even better speedup of $O(d +m)$ has been obtained with permutation-based LSH, such as minwise hashing, using ideas of densification~\cite{shrivastava2014densifying,shrivastava2014improved2,shrivastava2016simple,shrivastava2017Optimal}. These drastic reductions in hashing time have been instrumental in making LSH based algorithms more appealing and practical.

\section{Our Proposal}
\label{sec:proposal}

Denote the dataset with $\mathcal{D}=\{x_{(i)} | i \in [1,n]\}$, where $n$ is the number of data points in $\mathcal{D}$.  Unsupervised anomaly detection relies on some scoring mechanism, denote it by $S: \mathbb{R}^d \times \mathcal{D} \mapsto \mathbb{R}$, for every $x_i \in \mathcal{D}$, such that the value of $S(x_i,\mathcal{D})$ for $x_i \in \mathcal{D}$ is significantly different compared with  $x_i \in \mathcal{O}$. Here $\mathcal{O}$ is the outlier set.

By definition, outliers are significantly separated from an average data point. Therefore, any reasonable statistics of $x_i$ with respect to all other $x_j \in \mathcal{D}$ will be different for outliers compared to a normal data point. Even an average distance of $x_i$ with all other elements of $\mathcal{D}$ is a reasonably good statistics~\cite{ramaswamy2000efficient}. However, as noted before, computing these statistics requires storing the complete data $\mathcal{D}$. Note, that in general calculating every single $S(x_i,\mathcal{D})$ requires one complete pass over the dataset $\mathcal{D}$.   In addition, our experiments show that alternative estimations based on random sampling and random projections still lead to significant computational overheads.

We instead focus on classes of scoring functions $S(.,.)$ over the dataset that can be estimated efficiently using a tiny memory efficient data structure that can easily fit fast processor cache. Furthermore, we also want to update the data structure on the fly. In particular, any change in data from $\mathcal{D}$ to $\mathcal{D}'$ requires no change, and the estimates get dynamically adjusted.

We show that a class of scoring function of the following form has the required property:
\begin{equation}\label{eq:thescore}
    S(q,\mathcal{D}) = \sum_{x_i \in \mathcal{D}} p(q,x_i)^K,
\end{equation}
where $p$ is the collision probability of any LSH family and $K \ge 1$ is an integer.

The analysis of this paper extends naturally to any LSH scheme. For this work, we will focus on the popular signed random projections (SRP) as the LSH because of its simplicity. Furthermore, advances in fast SRP has lead to some very lightweight hashing variants. With SRP, the collision probability $p(q,x_i)$ is given by the formula:
\[
p(q,x_i) = 1 - \frac{1}{\pi}\cos^{-1} (\frac{q^T x_i}{\left \| q \right \| \left \| x_i \right \|})
\]
which will also be the value of $p(q,x_i)$ for the rest of the paper.

\subsection{Can it discriminate Outliers?}
\label{sec:whydiscrim}
To demonstrate the discriminative power of the scoring function in Equation~\ref{eq:thescore},  we do a simulation experiment similar to the one performed in~\cite{pham2012near}.

\begin{figure*}
	\centering
	\begin{subfigure}[b]{0.40\textwidth}
		\includegraphics[width=\textwidth]{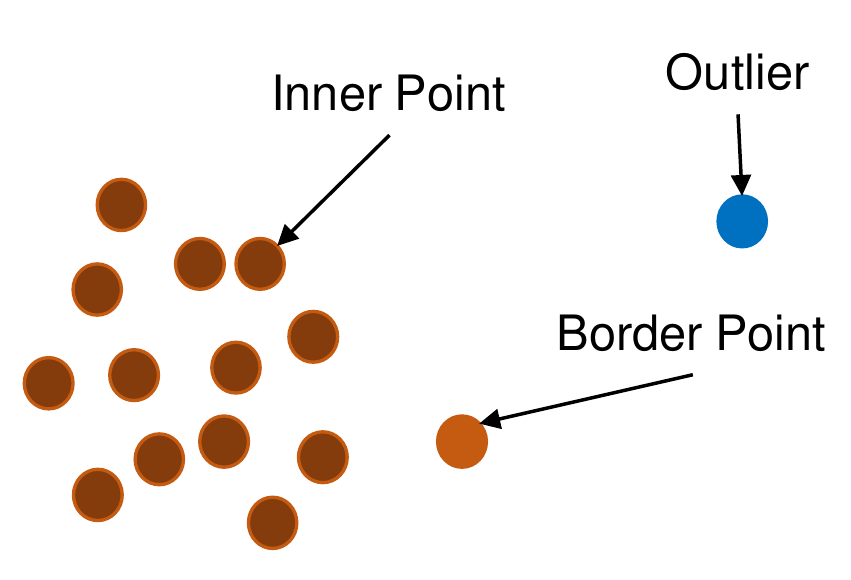}
		\caption{Illustration of simulated data with Inner Points, Border Points and Outliers.}
		\label{fig:exp1}
	\end{subfigure}
	~ 
	\begin{subfigure}[b]{0.50\textwidth}
		\includegraphics[width=\textwidth]{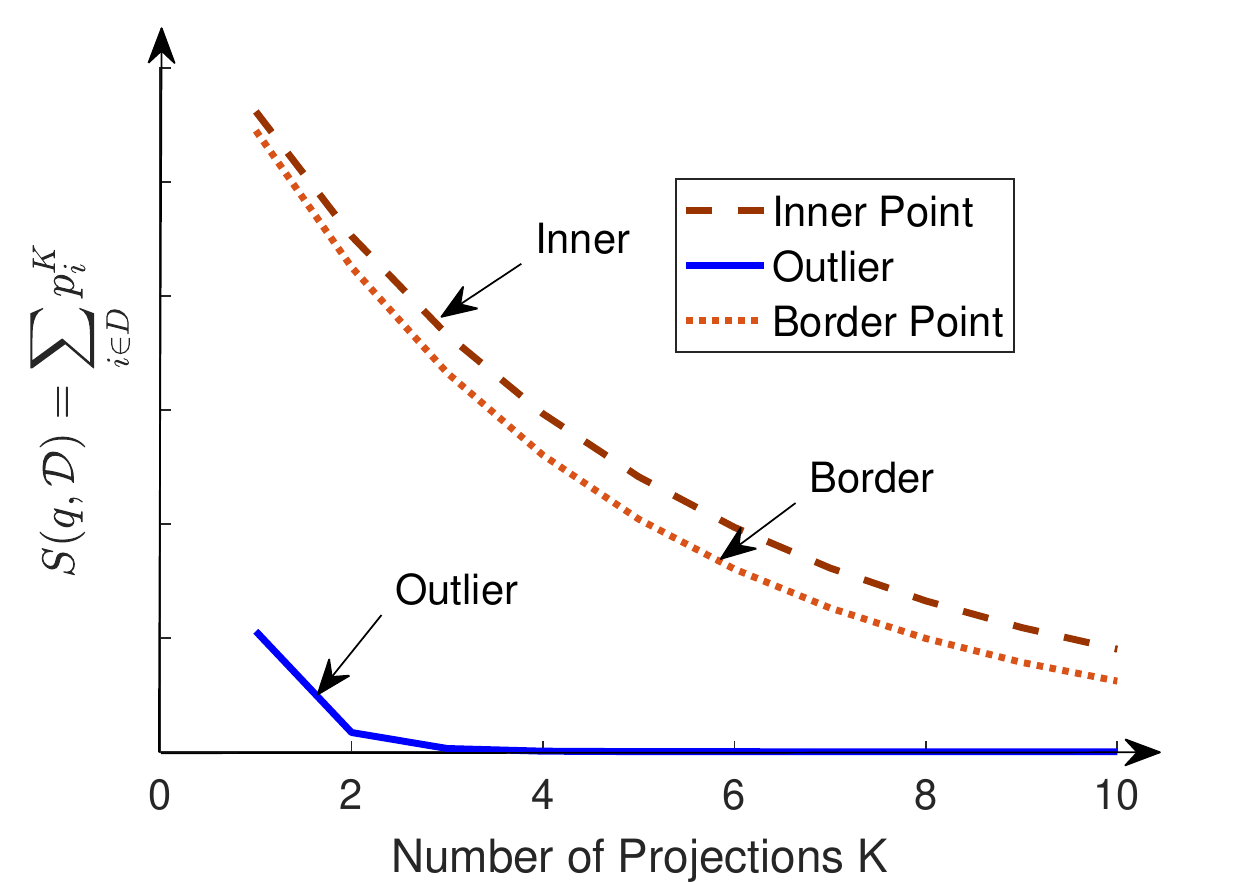}
		\caption{Normalized $S(q,\mathcal{D})$ with respect to $K$ for Inner Points, Border Points and Outliers.}
		\label{fig:exp2}
	\end{subfigure}
	\caption{ {\bf Discriminative power of $S(q,\mathcal{D})$:} We can see from the figure that the value of $S(q,\mathcal{D})$ for an Outlier is significantly lower (different) compared to that of non-outliers.}\label{fig:example}
\end{figure*}

We first generate a simple dataset with an outlier point.  Figure \ref{fig:exp1} shows the snapshot of the data. There are two sets of data points. The outlier and the general data points.  For the general data points, in addition, we make a distinction between the border points and inner points as illustrated in the figure.

In Figure \ref{fig:exp2}., we plot the of the value of $\frac{1}{n}S(q,\mathcal{D})$, given by Equation~\ref{eq:thescore}, for different sets of data points as a function of $K$.

We can see from the figure the value of our statistics $\frac{1}{n}S(q,\mathcal{D})$ for an outlier point is near zero. In particular, it is significantly lower compared to the values of the same statistics for inner points and even border points. This behavior is expected.
Note, that our statistics is a sum of collision probabilities of the LSH mapping over all the data points $x_i \in \mathcal{D}$. From the theory of LSH, the collision probability $p(q,x_i)$ indicates the level of similarity between $q$ and $x_i$. If $q$ is an outlier, $p(q,x_i)$ is expected to be significantly low.  We will further demonstrate the usefulness of this statistics in the experiments section.

\subsection{ACE (Arrays of (locality-sensitive) Counts Estimator) Algorithm}
\label{sec:ACEAlgo}
\begin{algorithm}[tb]
	\caption{\textbf{Arrays of (locality-sensitive) Count Estimator(ACE) Algorithm}}
	\label{alg:lsc}
	\begin{algorithmic}[1]
		\STATE {\bfseries Input:} Dataset $D$, Number of Hashes $K$, Number of Hash tables $L$, $\alpha$
        \STATE {\bfseries Hash Initialize:} Generate $L$ $H_j(.)$ using $K$ independent SRPs each.
        \FOR{$i=1$ {\bfseries to} $L$}
        \STATE $A_j = new \ short[2^K](0)$ (Short Arrays)
        \STATE $\mu = 0$, $n=0$
        \ENDFOR

        \hrulefill

        \vspace{0.1in}

        \STATE  {\bfseries Pre-Processing Phase}
        \FOR{$x_i \in \mathcal{D}$}
        \STATE $\mu_{incre} = 0$
        \FOR{$j=1$ {\bfseries to} $L$}
        \STATE   $A_j[H_j(x_i)]++$
        \STATE $\mu_{incre} = \mu_{incre} + \frac{2A_j[H_j(x)] + 1}{L}$
        \ENDFOR
        \STATE $\mu = \frac{1}{n+1}\bigg(n\mu + \mu_{incre}\bigg)$
        \STATE n++;
        \ENDFOR

        \hrulefill

        \vspace{0.2in}

        \STATE  {\bfseries Query Phase: Given query $q$}
        \STATE $\widehat{S(q,\mathcal{D})} = 0$
        \FOR{$j=1$ {\bfseries to} $L$}
        \STATE   $\widehat{S(q,\mathcal{D})} = \widehat{S(q,\mathcal{D})} + \frac{1}{L}A_j[H_j(x_i)]$
        \ENDFOR
        \IF{$\widehat{S(q,\mathcal{D})} \le \mu - \alpha$}
        \STATE report $q$
        \ENDIF
	\end{algorithmic}
\end{algorithm}

For the ease of explanation, we first describe the procedure of our proposed ACE algorithm. We later show that this procedure is an efficient statistical estimator of our proposed outlier score $S(q,\mathcal{D})$ defined by Equation~\ref{eq:thescore}.

The overall process of ACE is summarized in Algorithm~\ref{alg:lsc}. Our ACE algorithm, uses $K \times L$ independent SRP hash functions $h_i$, each given by Equation~\ref{eq:srp}. $K$ and $L$ are hyperparameters that are pre-specified. Note, this is analogous to the traditional $(K, L)$ parameterized LSH algorithm for near-neighbor search. However, we do not perform any retrieval which requires heavy hash tables with buckets of candidates for each hash index.  For near-neighbor, we further need to compute the distances of these candidates to identify the best.

Instead, our method only needs a check the value of a simple counter at each index. We only need arrays of counters. The process is significantly efficient, both in memory and speed, compared to a single LSH near-neighbor query.

\begin{figure}[t]
    \centering
    \includegraphics[width=0.5\textwidth]{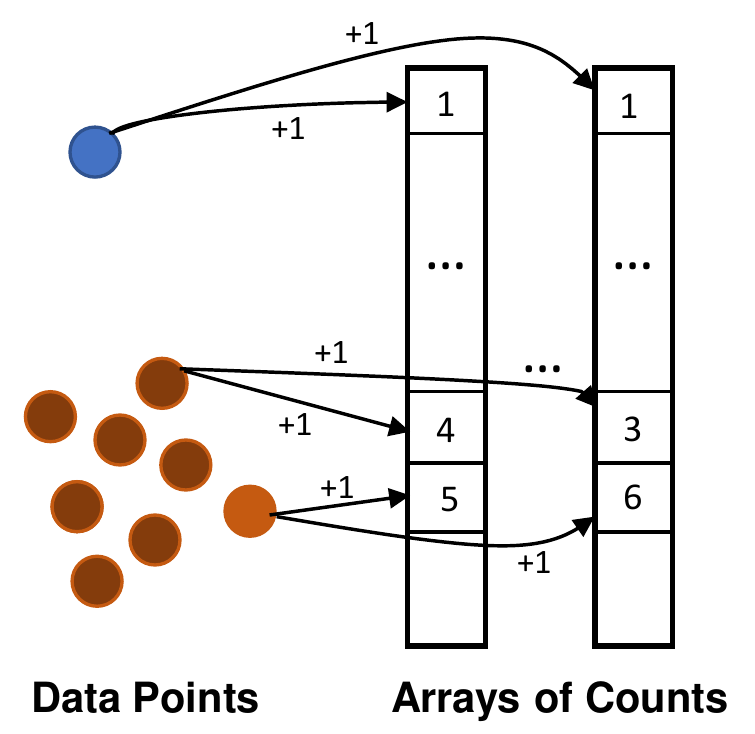}
    \label{fig:lookup}
    \caption{We use the LSH hash of the data points to increment corresponding counters into different (independent) hash arrays. We do not save anything, we only increase the value by $1$ for each bucket and then forget the data.}
\end{figure}

We use Signed Random Projections(SRP) $h^{sim}$ (Equations~\ref{eq:srp}) which gives one-bit output. Using these 1-bit outputs, we then generate $L$ different meta-hash functions given by  $H_j(x)  = [h_{j1}(x);h_{j2}(x);...;h_{{jK}}(x)]$ of $K$ bits each. The $K$ bits are generated by concatenating the individual bits. Here $h_{ij}, i \in \{1,2,...,K \}$ and $j \in  \{1,2,...,K \}$, are $K \times L$ independent evaluations of the SRP.

The overall algorithm works in the following two phases:\\
1) {\bf Counting Phase:} We construct $L$ short arrays, $A_j$, $j = \{1,\ 2, ..,\ L\}$, of size $2^{K}$ each initialized with zeros.  Given any observed element $x \in \mathcal{D}$, we increment the count of the corresponding counter $H_j(x)$ in array $A_j$, for all $j$s.  Thus, every counter keeps the total count of the number of hits to that particular index. The total cost of updating the data structure for any given $x$ is $KL$ SRP computations followed by $L$ increments.\\
{\bf Mean Update on Fly:} For each $x_i \in \mathcal{D}$ our estimated score is $$\widehat{S(x_i,\mathcal{D})} = \frac{1}{L}\sum_{j=1}^L A_j[H_j(x_i)].$$ We compute the mean behavior $\mu$ of the scores over all the element in dataset $x \in \mathcal{D}$.  $$\mu = \frac{1}{n}\sum_{i=1}^{n}\widehat{S(x_i,\mathcal{D})}.$$ Deviation from this mean will indicate outlierness. It turns out that we can dynamically update the mean $\mu$ on fly, as we read (or observe) the new data as shown in the Algorithm. See Section~\ref{sec:dynup} for details.\\
2) {\bf Real-time (query) Phase:} Given a query $q$, for which we want to compute the score, we report the average of all the counters $A_j[H_j(q)]$ $\forall j \in \{1,2,...,L\}$, i.e., $\widehat{S(q,\mathcal{D})} = \frac{1}{L}\sum_{j=1}^L A_j[H_j(q)]$. We report $q$ as anomaly if the estimated score $\widehat{S(q,\mathcal{D})}$ is less than $\mu - \alpha$, where $\alpha$ is some preselected hyperparameter. The overall cost for querying is $KL$ SRP computations and $L$ lookups followed by a simple average calculation.

\subsection{Theory: Analysis and Superiority over Random Sampling}
\label{sec:theory}
We first define few notations needed for analysis. Given a query point $q$.  For convenience, we will denote $p(q,x_i)$, the collision probability of the SRP of  $q$ with that of $x_i \in \mathcal{D}$, by $p_i$.

{\bf Intuition: LSH as Samplers} LSH is widely accepted as a black box algorithm for near neighbor search. We take an alternative adaptive sampling view of LSH which has emerged very recently~\cite{spring2017scalable,2017arXiv170305160S}. As argued in Section~\ref{sec:srp}, for a given query $q$ and $K$-bit SRP hash function $H_j$, the probability that any element $x_i$ increments the count of location $H_j(q)$ (the location of query) in array $A_j$ is precisely $p(q,x_i)^K$. Using this observation, we will show that the count of the number of elements, from $\mathcal{D}$, hitting the bucket of query $H_j(q)$ is an unbiased estimator of the $S(q,\mathcal{C})= \sum_{i=1}^n p(q,x_i)^K$. This novel use of LSH as efficient data structure for statistical estimation could be of independent interest in itself.

We define indicator variable $\mathbb{I}_{x_i \in B_q}$ as
\begin{align}\label{eq:indicator}
  \mathbb{I}_{x_i \in B_q} = \begin{cases}
                               1, & \mbox{if $x_i$ is in the bucket of $q$}  \\
                               0, & \mbox{otherwise}.
                             \end{cases}
\end{align}
That is $\mathbb{I}_{x_i \in B_q}$ is an indicator for the event that data element $x_i$ and the query $q$ are in the same bucket. It should be noted that
\begin{align}\label{eq:indicproperty}
  Pr(\mathbb{I}_{x_i \in B_q} =1) &= p(q,x_i)^K = p_i^K
\end{align}

Note that, $\mathbb{I}_{x_i \in B_q}$ and $\mathbb{I}_{x_j \in B_q}$ are correlated. If $x_i$ and $x_j$ are ``similar" then $\mathbb{I}_{x_i \in B_q}=1$ is likely to imply $\mathbb{I}_{x_j \in B_q}=1$. In other words, high similarity indicates positive correlation. Due to correlations, we may have both the cases:
\begin{align}
  \mathbb{E}[\mathbb{I}_{x_i \in B_q}\mathbb{I}_{x_j \in B_q}] &\begin{cases}
                                                             \ge p_i^Kp_j^K, & \mbox{(positive correlations)} \\
                                                            \le p_i^Kp_j^K, & \mbox{(negative correlation)}.
                                                           \end{cases}
\end{align}
Here, $\mathbb{E}$ is the expectation.

Using the above notations we can show that, for a given query $q$, $\widehat{S(q,\mathcal{D})}$, computed in Algorithm~\ref{alg:lsc}, is an unbiased estimator of $S(q,\mathcal{D})$ with variance given by:
\begin{theorem}
\label{theo:1}
  \begin{align}\notag
    \mathbb{E}[\widehat{S(q,\mathcal{D}}] &= \sum_{x_i \in \mathcal{D}} p_i^K = {S(q,\mathcal{D})} \\ \notag
    Var(\widehat{S(q,\mathcal{D})}) &= \frac{1}{L}\bigg(\sum_{i=1}^n  p_i^K(1 -  p_i^K) \\ \notag
    &+ \sum_{i \ne j} \big[\mathbb{E}[\mathbb{I}_{x_i \in B_q}\mathbb{I}_{x_j \in B_q}] -  p_i^K p_j^K\big]\bigg)
  \end{align}
\end{theorem}
The variance of $\widehat{S(q,\mathcal{D})}$ is dependent on the data distribution. There are two terms in the variance $\frac{1}{L}\bigg(\sum_{i=1}^n  p_i^K(1 -  p_i^K)\bigg)$ and $\frac{1}{L}\sum_{i \ne j}\big[\mathbb{E}[\mathbb{I}_{x_i \in B_q}\mathbb{I}_{x_j \in B_q}] -  p_i^K p_j^K\big]$. The terms inside summation is precisely the covariance between $\mathbb{I}_{x_i \in B_q}$ and $\mathbb{I}_{x_j \in B_q}$
\begin{align}\label{eq:Cov}
  \mathbb{E}[\mathbb{I}_{x_i \in B_q}\mathbb{I}_{x_j \in B_q}] -  p_i^K p_j^K &=  \mathbb{E}[\mathbb{I}_{x_i \in B_q}\mathbb{I}_{x_j \in B_q}]\\
  &- \mathbb{E}[\mathbb{I}_{x_i \in B_q}]\mathbb{E}[\mathbb{I}_{x_j \in B_q}] \\
  &= Cov(\mathbb{I}_{x_i \in B_q},\mathbb{I}_{x_j \in B_q})
\end{align}

There are $n(n-1)$ covariance terms in the second terms of variance, $\frac{1}{L}\sum_{i \ne j}\big[\mathbb{E}[\mathbb{I}_{x_i \in B_q}\mathbb{I}_{x_j \in B_q}] -  p_i^K p_j^K\big]$. To see why almost all of them will be negative, let $m$ be the number of elements in the buckets of the query. So only pairs $x_i$ and $x_j$ in the bucket ($O(m^2)$ pairs) of query $H_j(q)$ will contribute $1 - p_i^K p_j^K \ge 0$ to the summation (product of indicators is 1 $\iff$ both are 1). Rest all pairs ($O((n-m)^2)$) will contribute negative terms $-p_i^Kp_j^K$. Thus, if we choose $K$ large enough then the expected number of elements in the bucket $m$ is quite small. Hence, we can expect the variance to be significantly smaller than $\bigg(\sum_{i=1}^n  p_i^K\frac{(1 -  p_i^K)}{L}\bigg)$. We observe in our experiments that $K = 15$ is a good recommended constant value.

As noted the variance is dependent on the data distribution. If we have all exact duplicates, then all the covariances are positive. However, for real datasets, for any randomly chosen pair $x_i, \ x_j$, the covariance $Cov(\mathbb{I}_{x_i \in B_q},\mathbb{I}_{x_j \in B_q})$ will be almost always be negative.

An alternative way of estimating $S(q,\mathcal{D})$ is to use the random sampling. The idea is to uniformly sample a subset $\mathcal{S} \subseteq \mathcal{D}$ of size $L$ and report the random sampling estimator $RSE(q,\mathcal{D})$
\begin{align}\label{eq:RSE}
  RSE(q,\mathcal{D}) = \frac{n}{L} [\sum_{x_i \in \mathcal{S}}p_i^K]
\end{align}
From the theory of random sampling this estimator is also unbiased and has the following variance:
\begin{theorem}
\label{theo:2}
  \begin{align}\notag
    \mathbb{E}[RSE(q,\mathcal{D})] &= \sum_{x_i \in \mathcal{D}} p_i^K = {S(q,\mathcal{D})} \\ \notag
    Var(RSE(q,\mathcal{D})) &= \bigg(\frac{n}{L}-1\bigg)\bigg(\sum_{i=1}^n  p_i^{2K}\bigg)\\ \notag
    &= \sum_{i=1}^n  p_i^{K}\bigg(\bigg[\frac{n}{L}-1\bigg]p_i^K\bigg)
  \end{align}
\end{theorem}
Both $RSE(q,\mathcal{D})$ and $\widehat{S(q,\mathcal{D})}$ are unbiased. For the same number of samples, the estimator with smaller variance is superior.

We can get some insights from the leading terms $\sum_{i=1}^n  p_i^{K}\bigg(\bigg[\frac{n}{L}-1\bigg]p_i^K\bigg)$ and $\frac{1}{L}\bigg(\sum_{i=1}^n  p_i^K(1 -  p_i^K)\bigg)$. Generally, for any $i$ and large enough $n$, we always have $\bigg[\frac{n}{L}-1\bigg] \ge \frac{1}{L}(\frac{1-p_i^K}{p_i^K})$. Thus, for large enough $n$, $$Var(RSE(q,\mathcal{D})) > \frac{1}{L}\bigg(\sum_{i=1}^n  p_i^K(1 -  p_i^K)\bigg)$$
As argued before for real data we can expect $\frac{1}{L}\bigg(\sum_{i=1}^n  p_i^K(1 -  p_i^K)\bigg) > Var(\widehat{S(q,\mathcal{D})})$. Precise mathematical comparison between the variances of these two estimators is fairly challenging due to data-dependent correlation.

\begin{figure}[h]
	\centering
	\includegraphics[width=0.5\textwidth]{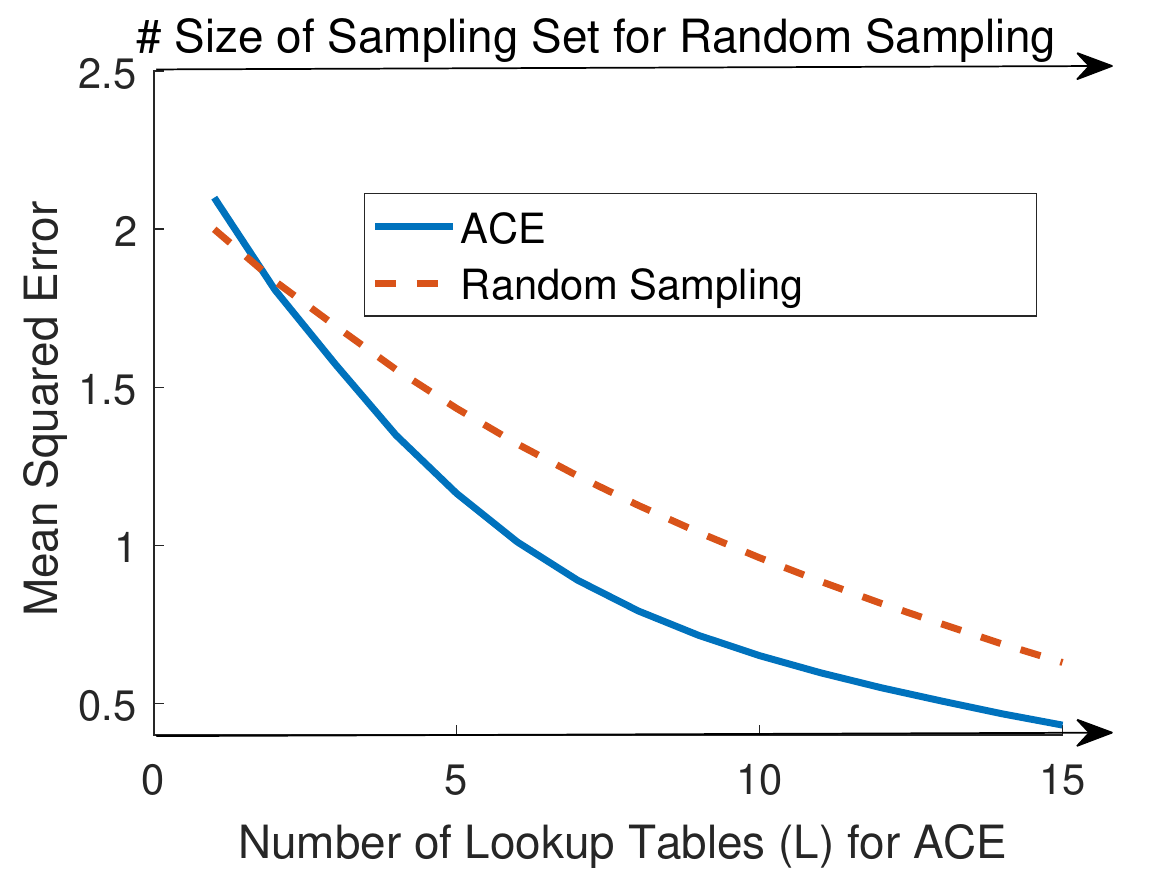}
	\caption{Comparison of ACE estimator with random sampling estimator on Image Object dataset, with varying $L$. ACE Estimator is consistently superior.}
\label{fig:compare}
\end{figure}
\begin{figure}[h]
	\centering
	\includegraphics[width=0.5\textwidth]{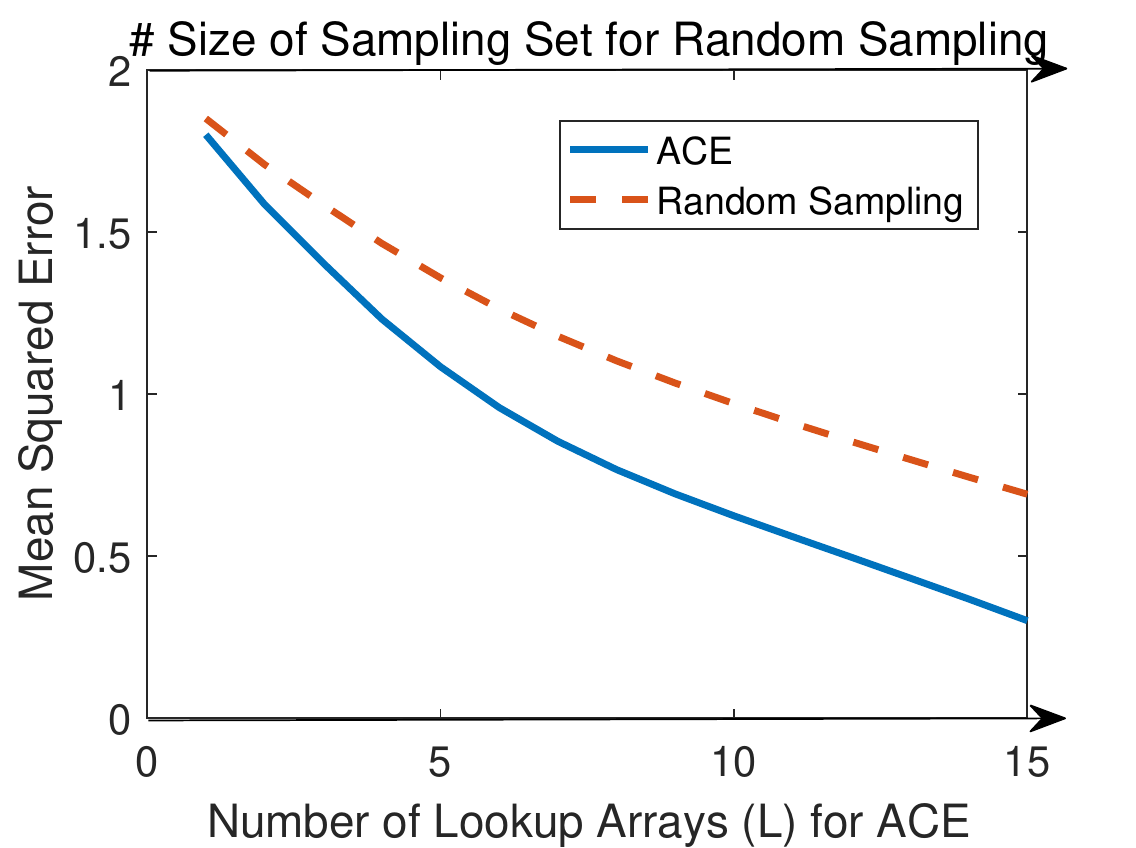}
	\caption{Comparison of ACE estimator with random sampling estimator on  KDDCUP-99 HTTP dataset, with varying $L$. ACE Estimator is consistently superior.}
\label{fig:comparekdd}
\end{figure}
\begin{figure}[h!]
	\centering
	\includegraphics[width=0.5\textwidth]{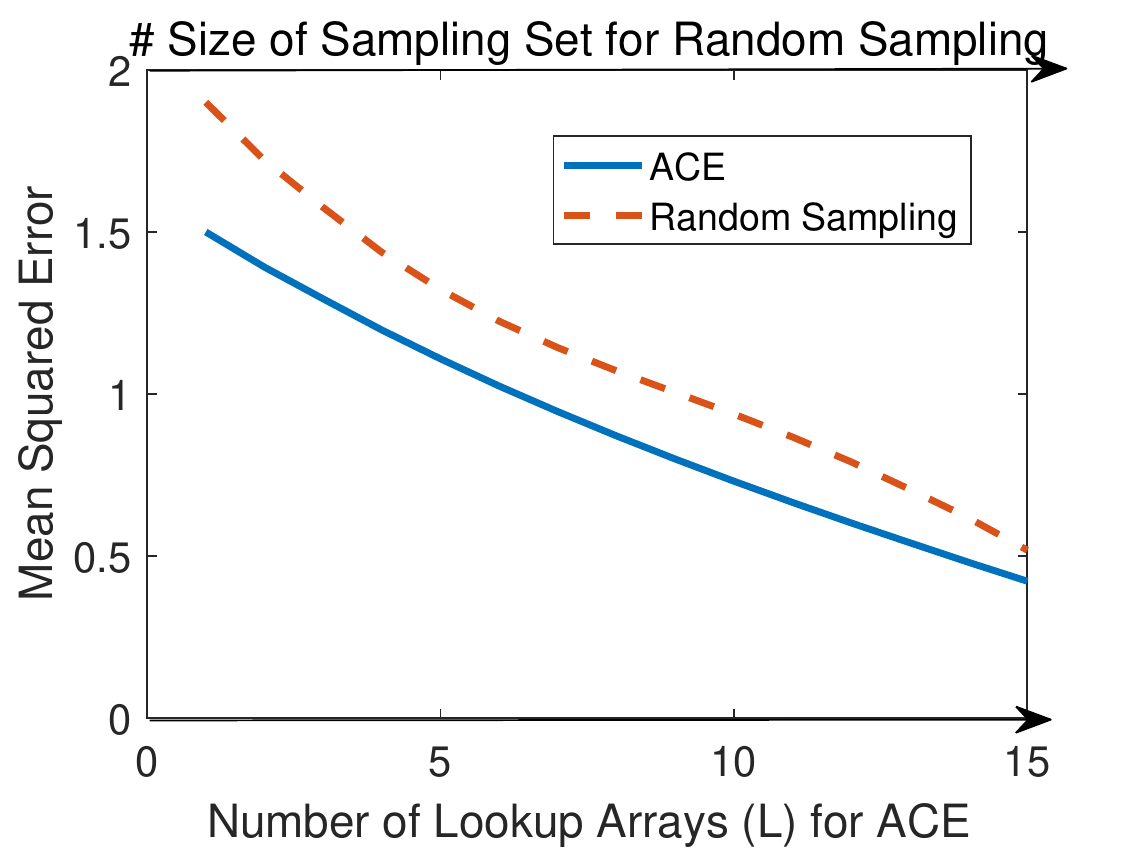}
	\caption{Comparison of ACE estimator with random sampling estimator on Statlog Shuttle dataset, with varying $L$. ACE Estimator is consistently superior.}
\label{fig:compareshuttle}
\end{figure}

{\bf Empirical Comparison:} As argued, we expect that for real datasets the ACE estimator to be more accurate (less variance) compared to the random sampling estimator. To validate our arguments empirically, we compare these estimators on the three benchmark anomaly detection datasets. These are the same datasets used in the experiments sections. See section~\ref{sec:datasets} for details. For all the three datasets, we randomly chose 50 queries and estimate their $S(q,\mathcal{D})$ using the two competing estimators. We use $K=15$ which is the fixed value used in all our experiments.

We plot the mean square error of the estimates, computed using the actual and the estimated values, in Figures~\ref{fig:compare},~\ref{fig:comparekdd},~\ref{fig:compareshuttle}. Different figures are for different datasets. We vary the number of samples for random sampling estimator $RSE(q,\mathcal{D})$ and the number of arrays for $\widehat{S(q,\mathcal{D})}$. From the plots, it is clear that on all the three real datasets, as expected from our analysis, our ACE estimator $\widehat{S(q,\mathcal{D})}$ consistently outperforms the random sampling estimator $RSE(q,\mathcal{D})$ at the same level of $L$. Note, these estimators are unbiased and hence mean square error value is also the theoretical variance. These experiments indicate that the variance of our ACE estimator is superior for estimating $S(q,\mathcal{D})$ over random sampling.

In addition to providing sharper estimates, in the next section, we show that our ACE algorithm only needs $O(d\log{d} + KL)$ computations to calculate the score. Here $d$ is the dimensions of the dataset.  For the same number of samples $L$, random sampling estimator requires $O(Ld)$ computations.  Given that $K=15$ is a fixed constant. For high dimensional datasets, we will have $d > K$. Thus, our estimator is not only more accurate but also cheaper compared to random sampling estimators from the computational perspective.

\subsection{Implementation Details, Running Time, Cache Utilization and Memory}
Computing the original score $S(q,\mathcal{D})$ requires $O(nd)$ computations and space, which for large and high dimensional datasets can be prohibitive.

{\bf Running Time:} From Algorithm~\ref{alg:lsc}, it is not difficult to see that for a query $q$, we need to compute $KL$ hashes of the data followed by a simple addition of size $L$. The costliest step is the computations of $KL$ hashes, which for $d$ dimensional data can be accomplished in $O(d\log{d} + KL)$ computations using advances in fast random projections (Section~\ref{sec:FastJL}). If instead, we are using minwise hashing as the LSH then it can be done in mere $O(d + KL)$ using fast minwise hashes. However, minwise hashing is limited to binary datasets only.

In all our experiments, we use $K=15$ and $L=50$ for all the three datasets irrespective of its size. Thus, with these small constant values, our scoring time negligible compared to the other algorithms which requires one pass over the full dataset $\mathcal{D}$.  In the experiments, we see that even with such minuscule computation, our method provides competitive accuracy while being orders of magnitude faster than 11 state-of-the-art methods.

{\bf Memory:}  Since we have $2^K$ counters, it is unlikely that the counters will get too many hits. To save memory by a factor of two, we can use short integers (16 bits) instead of integer counters. The total amount of memory required by $L$ counter arrays is $2^{K}\time 2$ bytes each if we use short counters. The total space needed for the arrays is $L \times 2^{K} \times 2$ bytes. For $K=15$ and $L=50$, the total space required by the ACE algorithm is around $3.2MB$. In addition, we need to compute $KL = 750$ hashes, which requires storing $750$ random seeds (integers) from which we can generate hashes on the fly. $750$ integers require negligible space compared to $3.2MB$. In the worst case, even if we decide to store the full random projections, we only need $750 \times d \times 8$ bytes (approx $6d$ kilobytes).

{\bf L3 Cache Utilizations:} For all of our experiments, the total memory requirement of the ACE algorithm is $\le 4MB$, irrespective of the size of the datasets. Our query data structure, the arrays, can easily fit into L3 cache of any modern processor, where the memory access can be anywhere from 2-10x faster than the main memory (DRAM) access. Detecting anomaly requires scoring which only needs reading count from the arrays. Due to all these unique favorable properties, our algorithm is orders of magnitude faster than the fastest available packages for unsupervised anomaly detection.

\subsubsection{Dynamic Updates}
\label{sec:dynup}
One of the appealing features of the ACE algorithm is that data can be dynamically updated and deleted. It is straightforward to increment (or decrement) the counters if we decide to add (or remove) any data $x$. However, we lose all the data information. We only store a set of count arrays, so it is not clear how we update the global mean $\mu$ of counts. Updating $\mu$ is an important part of Algorithm~\ref{alg:lsc}. Note, that the updated mean, $\mu'$, should be the average of all the estimated score of all the data in $\mathcal{D}'= \mathcal{D} + x$.

It turns out that we can exactly compute the new value of $\mu'$ from the existing count arrays. To simplify, let us convert old mean $\mu$ to sum by multiplying it by the size of dataset $n = |\mathcal{D}|$. It is easy to keep track of the sum $$n\mu = \sum_{x_i \in \mathcal{D}} \frac{1}{L}\sum_{j=1}^{L} A_j[H_j(x_i)].$$ Observe that, if the new $x$ goes to location $H_j(x)$ in array $A_j$ for any $j$. The count of location $H_j$ will be increment by 1. This will also lead to an increment in the scores of all the elements which maps to $H_j(x)$ in $j^{th}$ array by exactly $\frac{1}{L}$. Since we already know the count value of $A_j[H_j(x_i)]$, the total increment to the sum would be $\frac{A_j[H_j(x_i)]}{L}$. In addition, the new data $x$ will add an extra $\frac{A_j[H_j(x_i)]+1}{L}$ for its own count. Thus, we can precisely compute the increment in the sum. The new mean $\mu'$, for an addition of data $x$, can be computed as
\begin{align}\label{eq:dynadd}
   \mu' = \frac{1}{n+1}\bigg(n\mu + \sum_{j=1}^L \frac{2A_j[H_j(x)] + 1}{L}\bigg)
\end{align}
Similar arguments can be made for deletion of $x$, with deletion, the new mean is given by
\begin{align}\label{eq:dynsub}\mu' = \frac{1}{n-1}\bigg(n\mu - \sum_{j=1}^L \frac{2A_j[H_j(x)] - 1}{L}\bigg)\end{align}

\section{Discussions: Privacy Preserving Anomaly Detection}
Privacy is becoming one of the sought after directions in data mining and machine learning. Privacy preserving anomaly detection is of broad interest in the big-data and IoT (Internet of Things) community~\cite{vaidya2004privacy}. In many setting, we do want to detect anomalies in the data. However, it also desirable that the attribute information remains private and secure. It turns out that our proposed ACE algorithm has ideal properties for privacy preserving anomaly detection.

ACE does not require storing any data attributes, and the complete algorithm works only over aggregated counts generated from hashed data. If the hashes are not invertible, then the algorithm is safe. We can exploit advances in the secure computation to design protocols which hide the hashing mechanism~\cite{goldreich1998secure}.

Obtaining differential privacy~\cite{bhaduri2011privacy,machanavajjhala2008privacy} with ACE is quite appealing and neat. Since ACE algorithm relies on random projections to compute hashes, instead of original data, we can make ACE algorithm differentially private by adding only Gaussian noise instead of heavy-tailed Laplacian noise. ~\cite{kenthapadi2012privacy} shows a way to release user information in a privacy-preserving way for near-neighbor search. The paper showed that adding Gaussian noise $N(0, \sigma^2)$ after the random projection preserves differential privacy. Any function of differentially private object it also differentially private. Thus, to compute a private variant of SRP (Signed random projection), we used the sign of the differentially private random
projections (generated by adding Gaussian noise to usual projection) as suggested in~\cite{kenthapadi2012privacy}.

The final algorithm is very simple. The data is never revealed to anyone. At the source itself, the sign of differentially private random projections of data is used instead of usual SRP. All other process remains the same. Now since, we are only perturbing our algorithm with Gaussian noise, instead of Laplacian, we can expect a minimal loss in utility.

Note, that privacy is significantly harder with other state-of-the-art anomaly detection algorithms that store the actual data or even samples.  Making such algorithms private requires perturbing the system with heavy-tailed Laplacian noise, which can significantly hurt the outcome of the algorithm.

\section{Experimental Evaluations}
\label{sec:exp}
\subsection{Datasets}
\label{sec:datasets}We choose three real-world benchmark datasets for anomaly detection: 1) {\bf Statlog Shuttle}, 2) {\bf Object Images (ALOI)}, and 3) {\bf KDD-Cup99 HTTP}. These datasets are labeled and hence can be used for quantifying the effectiveness of anomaly detection measure. These three datasets also cover a broad spectrum of applications of unsupervised anomaly detection.

The first dataset we use is the shuttle dataset \footnote{https://archive.ics.uci.edu/ml/datasets/Statlog+(Shuttle)}.
This dataset describes radiator positions in a NASA space shuttle with 9 attributes.
It was designed for supervised anomaly detection. In the original datasets, about $20\%$ of the data regarded as anomaly. By following the preprocessing steps in \cite{abe2006outlier,reif2008anomaly}, we reduce the number of anomalies by selecting the class 1 as normal and apply a
stratified sampling for the classes 2, 3, 5, 6 and 7. The entire datasets contains $34,987$ instances with $879$ anomalies.

The second dataset is Object Images (ALOI) datasets\footnote{http://aloi.science.uva.nl/}.  The aloi dataset is derived from the “Amsterdam Library of Object Images” collection \cite{geusebroek2005amsterdam}. It contains about 110 images of 1000 small objects taken under different
light conditions and viewing angles. From the original images, a $27$ dimensional feature vector was extracted using HSB color histograms \cite{schubert2012evaluation}. Some objects were chosen as anomalies, and the data was down-sampled such that the resulting dataset contains $50,000$ instances including $1508$ anomalies.

The third dataset is KDD-Cup99 HTTP. \footnote{http://kdd.ics.uci.edu/databases/kddcup99/kddcup99.html}. KDD-Cup99 HTTP dataset \cite{leung2005unsupervised} is the largest benchmark for unsupervised anomaly detection evaluation. It contains simulated normal and attack traffic on an IP level in a computer network environment in order to test intrusion detection systems.  Following the preprocessing steps in \cite{leung2005unsupervised, carrasquilla2010benchmarking}, we use HTTP traffic only and also limit DoS traffic from the dataset.
Furthermore, the features of ``protocol" and ``port" information were removed. The remaining binary categorical features represented as 0 or 1 resulting in a total of $36$ dimensions. The dataset contains $596,853$ instances with $1055$ labeled anomalies.

The statistics of these datasets are shown in Table. \ref{tab:data}.

\begin{table*}[t]
	\centering
	\caption{The three datasets used in this paper with their statistics after standard preprocessing.}
	\label{tab:data}
	\begin{tabular}{c||c||c|| c}
		\hline \hline
		Dataset & Number of Instances & Number of Outliers & Dimension \\ \hline \hline
		Statlog Shuttle & $34,987$  & $879$ & $9$ \\ \hline
		Object Images (ALOI) & $50,000$ & $1508$ & $27$\\ \hline
		KDD-Cup99 & $596,853$ & $1055$ & $36$ \\ \hline \hline
	\end{tabular}
\end{table*}

\subsection{Baselines}
\label{sec:baseline}
\begin{table*}[h!]
	\centering
	\caption{Parameter settings for each algorithms on different dataset}
	\label{tab:para}
	\begin{tabular}{c||c||c||c}
		\hline
		\hline
		Method & Shuttle & Image Object & KDD-CUP 99 \\ \hline
		\hline
		ACE & $K=15, L=50$ & $K=15, L=50$ & $K=15, L=50$ \\ \hline
		LOF & $k=5 $ & $k=5$ & $k=10 $  \\ \hline
		kNN & $k=5 $ & $k=5 $ & $k=10 $  \\ \hline
		kNNW & $k=5 $ & $k=5 $ & $k=10 $  \\ \hline
		LoOP & $k_{reach} = k_{comp} = 5$  & $k_{reach} = k_{comp} = 5$ & $k_{reach} = k_{comp} = 10$  \\
		~ & $\lambda =0.2$  & $\lambda =0.2$ & $\lambda =0.2$  \\ \hline
		
		LDOF & $k=5$ & $k=5$ & $k=10$  \\ \hline
		ODIN & $k=5$ & $k=5$ & $k=10$  \\ \hline
		
		KDEOS & $k_{min} = k_{max} = 5$ & $k_{min} = k_{max} = 5$ & $k_{min} = k_{max} = 10$  \\
		~ & $Banwidth = 5, scale =0.2$ & $Banwidth = 5, scale =0.2$ & $Banwidth = 5, scale =0.2$\\
		~ & Gaussian Kernel & Gaussian Kernel & Gaussian Kernel\\ \hline
		COF & $k=5$ & $k=5$ & $k=10$  \\ \hline
		
		LDF & $h=1, c=0.1$ & $h=1, c=0.1$ & $h=1, c=0.1$  \\
		~ & Gaussian Kernel & Gaussian Kernel & Gaussian Kernel \\ \hline
		
		INFLO & $k=5, m=0.5$ & $k=5, m=0.5$ & $k=10, m=0.5$  \\ \hline
		
		FastVOA & $k=5$, $|S1| = 320$, $|S2| = 2$ & $k=5$, $|S1| = 320$, $|S2| = 2$ & $k=10$, $|S1| = 320$, $|S2| = 2$  \\ \hline
	\end{tabular}
\end{table*}
We use 11 different baselines methodologies to compare with ACE. These methodologies cover the whole spectrum of unsupervised anomaly detection techniques with all sorts of variations developed over the years. Our baselines cover scoring mechanisms based on simple to sophisticated strategies which include near-neighbor, kernel density estimation, graph connectedness, etc. First, we briefly describe all the competing algorithms:
\begin{enumerate}
  \item {\bf ACE:} This is the proposed Algorithm~\ref{alg:lsc}. We implement this algorithm in C++.
  \item {\bf LOF (Local Outlier Factor)~\cite{breunig2000lof}:} This is the most popular and the state-of-the-art unsupervised anomaly detection algorithm which use density-based local outlier factors in a database. LOF uses a score based on the difference between the local density of a point with that of its near-neighbors as the outlier score. LOF-ELKI is the fastest implementation of this algorithm.
  \item {\bf FastVOA (Fast Variance of Angles)~\cite{pham2012near}:} FastVOA is our benchmark sampling based methodology. FastVOA is a randomized algorithm which uses random sampling and random projections to estimate the variance of angle outlier measure. We use the C++ implementation\footnote{http://www.itu.dk/people/ndap/FastVOA.zip} provided by the authors.
  \item {\bf kNN (KNNOutlier)~\cite{ramaswamy2000efficient}:} This is the simplest method that uses the distance of an object to its $k$ nearest neighbor. We use the optimized implementation provided by the ELKI package.
  \item {\bf KNNW (KNNWeightOutlier) \cite{angiulli2005outlier}:} This is an improved version of the kNN algorithm where instead of the distance, we use the sum (accumulated) distance of a point to its $k$ nearest neighbors. Sum reduces the variance of the scores and it more stable. We again use the ELKI implementation.
  \item {\bf LoOP (Local Outlier probability) \cite{kriegel2009loop}:} This is a more advanced version of LOF. LoOP uses distance/density based algorithm similar to LOF to detect outliers, but with statistical methods to achieve better result stability. We use the implementation provided by ELKI.
  \item {\bf LDOF (Local Distance based Outlier Factor) \cite{zhang2009new}:} LDOF defines outliersness as the ratio of the average scores between the target point to all the other points in the kNN set, and of the sum all the pair-wise distance over all instances in the kNN set. Again we use the ELKI package.
  \item {\bf ODIN (Outlier Detection using Indegree Number) \cite{hautamaki2004outlier}:} ODIN defines outlierness as a low number of in-adjacent edges in the kNN graph. For more detail about this method, please refer \cite{hautamaki2004outlier}. ODIN is available in ELKI.
  \item {\bf LDF (Local density factor) \cite{latecki2007outlier} :} LDF replaces LOF's density function by a variance-width Kernel density estimation (KDE). In the KDE, the original (Euclidean) distance is replaced with the reachability distance of LOF. For more details of LDF, please refer \cite{latecki2007outlier}. LDF is available in ELKI.
  \item {\bf KDEOS (Kernel Density Estimation Outlier Score)} KDEOS also uses Kernel Density Estimation (KDE) in the LOF framework. KDEOS keeps the mathematical kernel density estimation intact for comparison with neighbor densities. The KDE densities are standardized per point as $z$-scores with respect to the KDE densities of the kNN set and averaged over different neighborhood sizes $k_{min}$... $k_{max}$. For more details of KDEOS, please refer to \cite{schubert2014generalized}.
  \item {\bf COF (Connectivity-based Outlier Factor) \cite{tang2002enhancing}:} COF modifies the density estimation of LOF to account for the ``connectedness" of a neighborhood via a minimum spanning tree (MST) rooted at the point under study. For details of COF please refer \cite{tang2002enhancing}.
  \item {\bf INFLO (Influenced Outlierness) \cite{jin2006ranking}:} INFLO compares the local model of LOF with the same density estimate applied to the reference set of the union of kNN and RkNN sets. INFLO is thus an example of a local outlier detection strategy for which different definitions of the neighborhood are used for the context set and reference set. For more details of INFLO, please refer \cite{jin2006ranking}.
\end{enumerate}

We use the highly optimized recent ELKI (Environment for Developing KDD-Applications Supported by Index-Structures) package\footnote{https://elki-project.github.io/} which is the most advanced set of anomaly detection algorithms noted for its efficient Java implementations. 10 of our baselines methodologies are implemented in this package. For FastVOA, a state-of-the-art randomized algorithm for variance of angle computation, we use the C++ package provided by the authors.

It should be noted that ACE and FastVOA are implemented in C++, while ELKI is a java package. A direct wall clock comparison is not fair. However, given the simplicity of our algorithm (Algorithm~\ref{alg:lsc}) which only requires simple hashing, use of primitive arrays, and simple summations. We do need any complex object other than arrays of short integers (primitives only). All other operations are primitive multiplications and summations. Thus, we expect that the difference between Java and C++ implementation would not be any significant for ACE. Furthermore, our results indicate a very significant speedup which cannot be explained by the difference in platforms.

\subsubsection{Parameter Settings}

Almost all of our baseline algorithms needed hyper-parameters. We use most of the default settings of the parameters as implemented. For baseline algorithms based on $k$ nearest neighbors, the ELKI package has the recommended settings for these benchmark datasets. To avoid complications, we directly use those recommended settings.  $k=5$ is recommended for the Image and the shuttle dataset while $k=10$ is recommended for the KDD-CUP dataset. We also observe that minor variations in $k$ do not lead to any significant changes. For kernel density estimation based algorithms, we use the default choice of Gaussian Kernel. For the sake to reproducibility, we summarize the precise hyperparameter settings of all the algorithms in Table~\ref{tab:para}. It should be noted that for ACE we use the fixed value of $K=15$ and $L=50$ for all the datasets. ACE does not need the near neighbor parameter $k$ (small)

\subsubsection{System and Platform Details}

We implemented our ACE algorithm in C++ and conducted experiments in a 3.50 GHz core Xeon Windows platform with 16GB of RAM.

We use g++ (version  5.4.0) as the C++ compiler for ACE and fastVOA. For running Java codes of ELKI package, we use OpenJDK 64bits version 1.8.0.

\subsection{Methodology and Results}


\begin{table*}[t]
	\centering
	\caption{Result on Statlog Shuttle Dataset}
	\label{tab:statlog}
	\begin{tabular}{c||c||c||c||c||c}
		\hline
		\hline
		& Outliers Reported & Correctly Reported & Outliers Missed & Execution Time (s) & Speed-up with ACE \\ \hline
		\hline
		ACE & 6763 & 273 & 606 & \textbf{0.81s} & ~1x\\ \hline
		LOF & 4356 & 381 & 498 & 14.12s & ~17.4x\\ \hline
		kNN & 4897 & 493 & 386 & 12.35s & ~15.2x\\ \hline
		kNNW & 5264 & 610 & 269 & 13.54s & ~16.7x\\ \hline
		LoOP & 6145 & 201 & 678 & 14.51s & ~17.9x\\ \hline
		LDOF & 6433 & 330 & 549 & 16.42s & ~20.3x\\ \hline
		ODIN & 9775 & 375 & 504 & 12.21s & ~15.1x\\ \hline
		KDEOS & 12630 & 314 & 565 & 11.73s & ~14.5x\\ \hline
		COF & 9133 & 280 & 599 & 13.45s & ~16.6x\\ \hline
		LDF & 9809 & 375 & 504 & 19.93s & ~24.6x\\ \hline
		INFLO & 4488 & 183 & 696 & 14.03 & ~17.3x\\ \hline
		FastVOA & 8532 & 271 & 608 & 235.10s & ~290.2x\\ \hline \hline
	\end{tabular}
\end{table*}

\begin{table*}[h!]
	\centering
	\caption{Result on Object Image Dataset}
	\label{tab:aloi}
	\begin{tabular}{c||c||c||c||c||c}
		\hline \hline
		& Outliers Reported & Correctly Reported & Outliers Missed & Execution Time (s) & Speed-up with ACE\\ \hline \hline
		ACE & 7216 & 340 & 1168 & \textbf{1.26s} & ~1x\\ \hline
		LOF & 4476 & 519 & 989 & 72.31s & ~57.4x\\ \hline
		kNN & 5428 & 447 & 1061 & 63.27s & ~50.2x\\ \hline
		kNNW & 5558 & 329 & 1508 & 89.96s & ~71.4x\\ \hline
		LoOP & 5121 & 253 & 1179 & 59.97s & ~47.6x\\ \hline
		LDOF & 7501 & 470 & 1038 & 60.39s & ~47.9x\\ \hline
		ODIN & 10110 & 162 & 1346 & 72.69s & ~57.6x\\ \hline
		KDEOS & 9515 & 404 & 1104 & 55.89s & ~44.36x\\ \hline
		COF & 8746 & 284 & 1224 & 81.74s & ~64.9x\\ \hline
		LDF & 9133 & 301 & 1207 & 60.51s & ~48.0x\\ \hline
		INFLO & 10328 & 420 & 1088 & 72.13s & ~57.2x\\ \hline
		FastVOA & 8931 & 319 & 1189 & 291.10s & ~231.0x\\ \hline \hline
	\end{tabular}
\end{table*}

\begin{table*}[h!]
	\centering
	\caption{Result on KDD-Cup99 HTTP Dataset}
	\label{tab:kddcup}
	\begin{tabular}{c||c||c||c||c||c}
		\hline \hline
		& Outliers Reported & Correctly Reported & Outliers Missed & Execution Time (s) & Speed-up with ACE\\ \hline \hline
		ACE & 22160 & 406 & 649 & \textbf{23.33s} & 1x\\ \hline
		LOF & 13260 & 523 & 532 & 1813.63s & ~77.7x\\ \hline
		kNN & 15432 & 365 & 690 & 1483.54s & ~63.5x\\ \hline
		kNNW & 14328 & 460 & 595 & 2125.43s & ~91.1x\\ \hline
		LoOP & 16578 & 396 & 659 & 1594.54s & ~68.3x\\ \hline
		LDOF & 16579 & 496 & 559 & 1674.43s & ~71.7x\\ \hline
		ODIN & 18054 & 365 & 690 & 1918.34s & ~82.2x\\ \hline
		KDEOS & 21095 & 469 & 586 & 1428.32s & ~61.2x\\ \hline
		COF & 20658 & 584 & 471 & 2043.43s & ~87.5x\\ \hline
		LDF & 19574 & 368 & 687 & 1485.85s & ~63.7x\\ \hline
		INFLO & 25704 & 565 & 490 & 1684.47s & ~72.2x\\ \hline
		FastVOA & 29316 & 354 & 701 & 3510.26s & ~150.4x\\ \hline\hline
	\end{tabular}
\end{table*}
All of these 12 algorithms associate a score with every element in the data. After association, a significantly lower score from the mean indicates an anomaly. In order to convert these scores into an anomaly detections algorithm, there are many reasonable strategies. We can rank each candidate, based on scores, and report bottom-k as the anomalies, but such rankings are not realistic. In real-time applications, we only see queries in an online fashion. Therefore, a more practical approach is to use a threshold strategy to report anomalies. We compute the mean $\mu$ and the standard deviations $\sigma$ of the scores on the dataset of interest and report any element with the associated score less than $\mu - \sigma$ as an anomaly.

With this anomaly detection strategy, we run all the 12 algorithms on the chosen three datasets. We report five different numbers separately for each of the three datasets: (1) Number of outliers reported, (2) Number of outliers correctly reported, (3) Number of Outliers missed, (4) the CPU execution time for the different methods, and (5) Relative speed with ACE. The CPU executing time does not include the end to end time of the complete run of the algorithm, which includes data reading, preprocessing (if any), scoring and outlier reporting. Relative speedup reports the ratio of the time required by a given algorithm to the time required by ACE algorithm.

The results are shown in Table \ref{tab:statlog}, Table. \ref{tab:aloi}, and Table \ref{tab:kddcup} for { Statlog Shuttle}, { Object Images (ALOI)}, and { KDD-Cup99 HTTP} datasets respectively.

Although, there is no clear winner in terms of accuracy. LOF seems to be consistently more accurate than others. The number of anomalies reported correctly (true positives) with ACE is similar to other algorithms. ACE, however, seems to report slightly more anomalies (high false positives) than other algorithms. This is not a major concern though. Few extra false positives are easy to deal with because we can always further filter them using a more sophisticated algorithm, so long as they are small. Overall, our proposed new scoring scheme $S(q,\mathcal{D})$ and the corresponding estimator performs very competitively, in terms of accuracy, in comparison with many successful algorithms.

The most exciting part is the computational savings with ACE. What we observe is that our methodology is significantly faster than any other alternative irrespective of the choice of dataset. ACE algorithms is at least around 15x, 45x and 60x faster than the best competitor on { Statlog Shuttle}, { Object Images (ALOI)}, and { KDD-Cup99 HTTP} datasets respectively. Most of the algorithms, based on near-neighbors except FastVOA, have similar speeds. This could be because almost all of them requires computation of the order of the data. FastVOA is consistently very slow, which we suspect is because the estimators used in FastVOA is computationally very expensive. FastVOA estimators require multiple sorting and frequently computing costly medians. See~\cite{pham2012near} for details. ACE is around 150-300x faster than FastVOA.

The results are even more exciting if we start considering the memory requirements. With $K=15$ and $L=50$, our methodology requires less than $4MB$ of operating memory for the complete run of the algorithm. Since we use the same $K$ and $L$ across all datasets, this $4MB$ requirement is unaltered. We never keep any data in the memory. On the other hand, all other methods except FastVOA require storing complete data in the memory. In our case, the KDD-Cup99 HTTP dataset itself is around 165MB to store. Although KDD-Cup99 HTTP dataset is the largest labeled benchmark, it is still tiny from big-data perspective.

For real large-scale streaming application, the data size can quickly go into terabytes. At such scales algorithms requiring storing and/or processing the complete data are infeasible.

The disruptive performance of ACE is not surprising given the simplicity of the process. However, as argued, the process is a statistically sound procedure for estimating the proposed score $S(q,\mathcal{D})$.

\section{Conclusion}

Statistical measures for popular learning and data mining problems, such as anomaly detection, were designed without taking into account the computational complexity of the estimation process. When faced with current big-data challenges, most of these estimation process fail to address tight resources constraints.  In this paper, we showed that for the problem of unsupervised anomaly detection, we could leverage advances in probabilistic indexing and redesign a super fast statistical measure which requires significantly lesser resources.

We proposed ACE algorithm, for unsupervised anomaly detection, which is 60-300x faster than existing approaches with competing accuracy. Our algorithm requires mere $4MB$ of memory which can utilize L3 caches of modern processors leading to around 2-10x savings in latency. The ACE algorithm can quickly adapt to drifting volumes of data and has very appealing privacy properties. We believe ACE will replace existing unsupervised anomaly detection algorithms deployed in high-speed high-volume data processing systems.




\newpage

\section*{Acknowledgments}
This work was supported by National Science Foundation (NSF) Award IIS-1652131.

\begin{appendix}
\section{Proof of Theorem 1}

Since every arrays are independent, due to independence of hash functions, the estimator $\widehat{S(q,\mathcal{D})}$ is an average of $L$ simple estimators. Consider, just the first array $A_1$ and the index corresponding to $q$ in this array is $H_1(q)$. Now the value of $A_1(H_1(q))$ can be written using the indicator variables $\mathbb{I}_{x_i \in B_q}$ (Equations~\ref{eq:indicator}) as:
\begin{align}
  A_1(H_1(q)) &= \sum_{i = 1}^{n} \mathbb{I}_{x_i \in B_q}.
\end{align}
Taking expectation on both side and noting the value of $\mathbb{E}(\mathbb{I}_{x_i \in B_q})$ from Equation~\ref{eq:indicproperty}, we get
\begin{align}
\mathbb{E}[A_1(H_1(q))] &= \sum_{i = 1}^{n} \mathbb{E}[\mathbb{I}_{x_i \in B_q}]\\
&= \sum_{i = 1}^{n} p_i^K = S(q,\mathbb{D})
\end{align}
Note, the linearity of expectation which is still valid even if the indicator variables are correlated.
Thus, the count inside $A_1(H_1(q))$ is an unbiased estimator of the score. Since our final estimator $\widehat{S(q,\mathcal{D})}$ is average of $L$ unbiased estimators $A_j(H_j(q))$, for $j =\{1, \ 2, ..., \ L\}$. The unbiasedness of $\widehat{S(q,\mathcal{D})}$ follows from the linearity of expectation.

Similarly for variance, we analyze the variance of $A_1(H_1(q))$,  $\widehat{S(q,\mathcal{D})}$ being average over $L$ independent estimators will have the variance $$var(\widehat{S(q,\mathcal{D}))} = \frac{1}{L}Var(A_1(H_1(q))).$$ The independence is important.
Again we have
\begin{align}
 A_1(H_1(q)) &= \sum_{x_i \in \mathcal{D}} \mathbb{I}_{x_i \in B_q}. \\
  Var(A_1(H_1(q))) &= \mathbb{E}(A_1(H_1(q))^2) - (\mathbb{E}(A_1(H_1(q))))^2  \\
  \mathbb{E}(A_1(H_1(q))^2) &= \mathbb{E}\bigg[\big(\sum_{x_i \in \mathcal{D}} \mathbb{I}_{x_i \in B_q}\big)^2\bigg] \\
   &= \mathbb{E}\bigg[\big(\sum_{x_i \in \mathcal{D}} \mathbb{I}_{x_i \in B_q}^2\big)\bigg]\\
   & +  \mathbb{E}\bigg[\sum_{x_i, \ x_j \in \mathcal{D}; i\ne j}\mathbb{I}_{x_i \in B_q}\mathbb{I}_{x_j \in B_q} \bigg] \\
   &= \sum_{i = 1}^{n}p_i^K  + \sum_{x_i, \ x_j \in \mathcal{D}; i\ne j}\mathbb{E}[\mathbb{I}_{x_i \in B_q}\mathbb{I}_{x_j \in B_q}]
\end{align}
Note $\mathbb{E}[\mathbb{I}_{x_i \in B_q}\mathbb{I}_{x_j \in B_q}]$ cannot be factored as indicators are not independent.

The final expression follows from substituting the following for $(\mathbb{E}(A_1(H_1(q))))^2$ and algebraic manipulations.
\begin{align}
  (\mathbb{E}(A_1(H_1(q))))^2 &= (\sum_{x_i \in \mathcal{D}} p_i^K)^2 \\
   &= \sum_{x_i \in \mathcal{D}}p_i^{2K} + \sum_{x_i, \ x_j \in \mathcal{D}; i\ne j}p_i^{K}p_j^K
\end{align}

\section{Proof of Theorem 2}

The random sampling estimator can be analysed by defining indicator variable $\mathbb{I}_{x_i \in \mathcal{S}}$ as
\begin{align}
  \mathbb{I}_{x_i \in \mathcal{S}} = \begin{cases}
                               1, & \mbox{if $x_i$ is sampled}  \\
                               0, & \mbox{otherwise}.
                             \end{cases}
\end{align}
Due to uniform sampling we have
\begin{align}
  \mathbb{E}[\mathbb{I}_{x_i \in \mathcal{S}}] = \frac{|\mathcal{S}|}{|\mathcal{D}|} = \frac{L}{n}
\end{align}
Using this indicator variable we can write the random sampling estimator as:
\begin{align}
  RSE(q,\mathcal{D}) &= \frac{n}{L}\sum_{x_i \in \mathcal{S}}p_i^K \\
   &=  \frac{n}{L}\sum_{x_i \in \mathcal{D}}  \mathbb{I}_{x_i \in \mathcal{S}} p_i^K
\end{align}
Taking expectation and utilizing linearity of expectation gives us the desired unbiasedness result.

The variance part is also straightforward algebraic expansion, similar to the proof of theorem~\ref{theo:1}. It leads to
\begin{align}
  \mathbb{E}[RSE(q,\mathcal{D})^2] &= \frac{n^2}{L^2}\bigg(\sum_{x_i \in \mathcal{D}} \mathbb{E}[\mathbb{I}_{x_i \in \mathcal{S}}] p_i^{2K}  \\
   & + \sum_{x_i,x_j \in \mathcal{D}; i\ne j} \mathbb{E}[\mathbb{I}_{x_i \in \mathcal{S}}\mathbb{I}_{x_j \in \mathcal{S}}] p_i^Kp_j^K\bigg)
\end{align}
Since $x_i$ and $x_j$ are sampled independently, $\mathbb{I}_{x_i \in \mathcal{S}}$ and $\mathbb{I}_{x_j \in \mathcal{S}}$ are independent for any $i$ and $j$. Therefore,
\begin{align}
  \mathbb{E}[RSE(q,\mathcal{D})^2]  & =\frac{n^2}{L^2}\bigg(\sum_{x_i \in \mathcal{D}} \frac{L}{n}p_i^{2K}\\
  &+  \sum_{x_i,x_j \in \mathcal{D}; i\ne j} \frac{L^2}{n^2}p_i^{K}p_j^{K} \bigg)
\end{align}
Subtracting the value of $\mathbb{E}[RSE(q,\mathcal{D})]^2$ further simplifies the variance to the desired expression.

\end{appendix}

\balance

\newpage
\bibliographystyle{abbrv}
\bibliography{lsh_anomaly}  

\end{document}